# Perspectival Computational Thinking for Learning Physics: A Case Study of Collaborative Agent-based Modeling

Amy Voss Farris & Pratim Sengupta
Mind, Matter & Media Lab, Vanderbilt University – Peabody College
Email: amy.farris@vanderbilt.edu, pratim.sengupta@vanderbilt.edu

**Abstract:** We examine the process through which computational thinking develops in a perspectival fashion as two middle school students collaborate with each other in order to develop computational models of two graphs of motion. We present an interaction analysis of the students' discourse and computational modeling, and analyze how they came to a joint understanding of the goal of the modeling activity. We show that this process involves bringing about coherence between multiple perspectives: the object in motion, the computational agent, the other student, and graphs of motion.

## Introduction & Background

Computational thinking, modeling and programming are now regarded as core epistemic and representational practices in K12 science and engineering (NGSS, 2013). Wing (2011) described computational thinking (CT) as the "thought processes involved in formulating problems and their solutions so that the solutions are represented in a form that can be effectively carried out by an information-processing agent" (p. 1). The development of computational thinking is manifested in and deeply intertwined with representational practices such as programming and modeling, which in turn involve the design and development of computational abstractions such as defining patterns, generalizing from instances, and parameterization (Wing, 2006; Sengupta et al., 2013). In this paper, we present a theoretical framework for analyzing and understanding the role of perspectives, or points of view, in the development of collaborative computational thinking in the context of using agent-based programming and modeling for learning physics.

Prior research on learning physics using collaborative agent-based modeling has identified how shared understanding among student dyads develops through divergence and convergence of their conceptual understandings (Roschelle, 1992; Roschelle & Teasley, 1995). The students in these studies used a modeling environment called the Envisioning Machine, in which they directly manipulated a graphical simulation of the velocity and acceleration of a computational agent (similar to a Logo turtle) by altering the settings of the velocity and the acceleration vectors. Similar to Roschelle and Teasley, we also use an agent-based modeling environment, because previous studies have shown that students can indeed develop deep understandings of motion as a process of continuous change using such learning environments (Sengupta & Farris, 2012). However, in our study, the use of programming plays a central role in the students' interactions.

## Theoretical Framework

At the broadest level, we seek to answer the following question: How does computational thinking begin to develop in a collaborative setting when students engage in agent-based programming in order to model the motion of an object as a process of change in distance and speed over time? To answer this question, we adopt the lens of *perspectival understanding* (Greeno & van de Sande, 2007; Greeno & MacWhinney, 2006). Greeno and van de Sande (2007) defined the construction of perspectival understanding as a process of constraint satisfaction. Following Thagard, they hypothesized that *coherence*—i.e., the compatibility or consistency of interacting representational elements (e.g., propositions, perspectives, meanings, etc.)—is the most general constraint (Thagard & Verbeurgt, 1998). Based on Greeno and his colleagues, we adopt the view that in a collaborative setting, manifold elements of the participating individuals' conceptions are taken up in a new joint understanding that is shaped by participants' points of view (POVs). For example, an interactant may operate from a point of view (POV) that is either enmeshed in the phenomenon or takes a depictive perspective, which views the phenomenon from a top-down or extrinsic perspective; however, the joint understanding that emerges during interaction can include expressed constituents of multiple POVs (Greeno & van de Sande, 2007).

Pedagogically, the ability of a user to take on the perspective of an actor within the system is an important affordance of agent-based models (Wilensky & Reisman, 2006), and virtual environments (Lindgren, 2012). In the context of understanding scientific phenomena that result from the aggregation of individual-level actions (*e.g.,* change in motion over time, ecological interdependence, or formation of traffic jams) adopting an agent-perspective enables the learner to use their intuitive knowledge—which is often in the form of embodied knowledge (Papert, 1980)—in order to develop a deep understanding of how the collective phenomenon emerges from the aggregation of individual, agent-level actions. However, there is little understanding of the process through which learners *begin* to develop a perspectival stance. Our study offers a window into this process. We show how two collaborating learners shift across and negotiate multiple perspectives—the object in motion, the other student, the programmable agent, and an aggregate, descriptive view of the graph—in order to interpret and model computationally the motion implied by the graph(s). In Thagard's terms, these perspectives represent the *elements* across which *coherence* is established through a process of shifts and negotiations between these perspectives (Thagard & Verbeurgt, 1998).

## Data Collection and Analysis

The study took place in a classroom at a large private university in the mid-southern USA. The topic of the course was scientific modeling, and the course met on the mornings of six consecutive Saturdays during the regular academic year. Twenty students were recruited via web posting on the university website, ages 10 - 12 (Grades 5 and 6). They were enrolled on a first-come, first-served basis and had no prior programming experience. The second author taught the course, while the first author acted as a facilitator during the third day of the course, when the activity we report here took place. The children in the focal dyad are two 10-year-old males, Arnav and Liam (pseudonyms), who volunteered to do the task as a pair and agreed to talk about their thinking with each other while they worked.

We present in-depth analysis of 23 minutes of collaboration between the students. We use the case study method (Yin, 2009) and interaction analysis methods (Jordan & Henderson, 1995). The entire activity was video recorded by two cameras focused on the children and their screens. Additional data were collected as saved files and screen captures from the student computers. We created multimodal transcripts of the discourse, and coded them line-by-line for the perspective, or POV, of the speaker. Our analysis focuses primarily on the students' talk, and teaching moves and interaction between researchers and students are also described where relevant. We describe three significant episodes of interaction, and within each episode, we present an analysis of a salient, smaller segment of discourse. We also present a brief, summative analysis of the shifts and coherence in perspectives during the entire interaction.

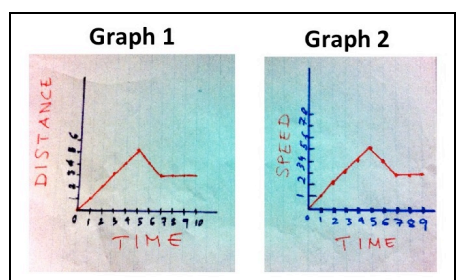

Figure 1. Distance-time (Graph 1) and speed-time graphs (Graph 2)

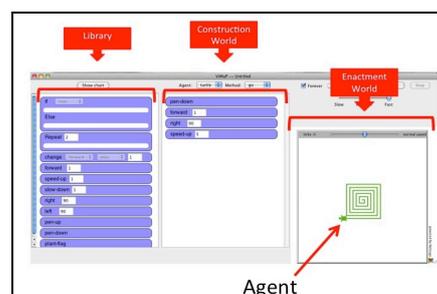

Figure 2. The ViMAP interface

## The Activity and Instructional History

The goal of the focal activity was to generate an agent-based computational model for each of the two graphs, distance versus time (Graph 1, Figure 1), and speed versus time (Graph 2, Figure 1). In each model, students were asked to represent the motion of an agent in a manner that would match the motion represented in the respective graph. The graphs, as shown in Figure 1, were digitally projected on a large screen in the front of the classroom for the duration of the activity. The students were working with an agent-based modeling and visual programming environment called ViMAP (Sengupta, Farris, & Wright, 2012), where the user creates a computer program by dragging and spatially arranging programming commands from a library of command blocks (Figure 2). ViMAP commands were specifically designed to support domain-specific learning in kinematics (Sengupta et al., 2012). When the user runs the program, these commands are then acted out in representational space by a Logo turtle in a NetLogo microworld (Wilensky, 1999). In the activities leading up to the episodes reported here, students had developed ViMAP models of motion in which they generated turtle geometry shapes, in which step-size of the turtle represented the speed of motion. For example, the rectangular spiral in Figure 2 represents constant acceleration, as the step size of the turtle increases by the same amount every turn. In these activities, students developed models of several different kinematic phenomena, including motion on a roller coaster. All the students were familiar with line graphs from their regular science and math classes, although the activity we report here was the first time they encountered graphs of motion during the study.

## Episode 1: Negotiating the Meaning of the Distance Graph

In the first episode Arnav and Liam engaged in negotiation of how to interpret Graph 1 (Figure 1). Arnav wanted to program the ViMAP turtle to reproduce the shape of the graph, which is a common novice approach to interpreting and representing graphs of motion (McDermott et al, 1987). Liam, however, argued that the graph meant that the distance was gradually going up. The segment below [2:29 – 2:48] illustrates Episode 1.

Arnav: Yeah, but no. Distance 6 ((left arm used to approximate slope of first segment of graph)).
Liam: It's GRADUALLY going up.
Arnav: No, because the next one is speed.
Liam: No, no, dude. Think about the first graph ((points to graph)). It's not you-something goes to 6, it's over time we go UP to 6. So then, (both?) is what we need to do.

Arnav's interpretation of Graph 1 ignores the role of time over which the change in distance takes place. We call this an *extrinsic graph* perspective, as he focuses on the shape of the line in Graph 1, and wants to program the ViMAP turtle to reproduce this line. In other words, even though he did consider the actions of the computational agent, these actions would simply reproduce the shape of the line in Graph 1—i.e., the graph extrinsic perspective was driving the actions of the agent. Liam countered with a more sophisticated interpretation: that the graph meant that something is "gradually going up" to 6. Liam interpreted the graph from an *intrinsic graph* perspective. That is, he focused on how the shape of the graph came to be, and to do so, he took on the perspective of actors ("we"), whose motion would gradually generate the shape of the graph ("...over time we go up to 6"). Also of interest here is that Arnav was unwilling to consider gradual change in distance as a representation of Graph 1, because Graph 2 shows speed—suggesting that he confused change in speed with change in distance over time.

## Episode 2: How Far from Where You are You're From

This episode began when Arnav and Liam asked for help from the first author (Amy), who in turn, asked them to explain to her what they inferred about the motion of the putative object from Graph 1 (Figure 1). The students described an object speeding up, slowing down, then staying the same. This verbal description matched their computational model, which used speed up and slow down commands. But the students were confusing speed with displacement; so, in order to encourage the students to think about displacement instead of speed, Amy then asked them if there are any two parts of the graph that are "the same." Arnav said that times 3 and 7 (on the y-axis) indicated the same "distance from the starting point," using the same words the instructor used with the whole class. After Amy left, the following exchange ensued [18:19 – 18:30], in which Arnav explains his interpretation of "distance from the starting point" to Liam.

Arnav: It's not depending on how LONG it is ((uses pen as pointer to make an invisible line across the table surface)), it's depending on how far from where you are you're from, not how long the roller coaster actually is.
Liam: Like he said, it how far FROM=
Arnav: =No, so, if you're here ((right hand in front of chin))… and then you do a loop ((half-circle upward motion)), and you come back ((half-circle downward motion)), you'll be pretty much at the same distance as you started from.

In Arnav's final turn of talk, he uses the example of a loop, similar to a vertical roundabout segment during a roller coaster ride. His embodied definition and gestural enactment evidence his shift to thinking about the motion from the perspective of the agent in motion, which is at the same position in the beginning (time = 3) and end (time = 7) of the loop. Arnav and Liam's ViMAP model represented the changes in displacement over time, which evidences an agent perspective of getting further from, then closer to, a point in space. Arnav's construction fused inanimate physics entities with flexibly construed animate objects. The verb forms are enduring and simple present. Taking the perspective of a non-human agent allows the embodied action to transcend time and setting. Throughout this segment, Arnav's egocentric perspective was merged with the object perspective. Arnav and the (imagined) object in motion are conjoined in simultaneous, multiple constructed worlds: the here and now of the interaction, the visual representation, and the imagined physical processes. Ochs and colleagues have identified this kind of speech in the discourse of professional physicists (Ochs et al., 1996). In this utterance, the physics entities are distances and points, articulated with "how long it is" (where 'it' refers to the length of the roller coaster track), and "how far from where you are you're from," in which "where you're from" is a point, and "how far you are from where you're from" is a distance. The only verb used in the utterance, "are," is simple present and enduring (Ochs et al., 1996), as it allows the embodied action to transcend time and setting. Similar indeterminate grammatical constructions, along with gestural journeys through visual displays, as observed in physicists' discourse, constitute physicist and physical entity as co-experiencers of dynamic processes (Ochs et al., 1996).

## Episode 3: Coordinating relationships among speed, distance, and time

In this episode, Arnav and Liam revised the loop to first make the turtle go down, and then up. Their thinking was that this motion could explain *both* Graph 1 and Graph 2, as the roller coaster car would get faster as it went down the downward half of the loop, and slow down as it went up the other half of the loop. The segment of conservation we report here [22:19 – 22:46] results from their efforts to repair trouble: the period of acceleration in Arnav's proposed model did not correspond with the period of "getting further away" in Graph 1.

Arnav: So we have to put it for 6 seconds ((adds repeating loop with a parameter of 6 six seconds))…forward 1, speed up 1 ((adds a forward command in the loop))
Liam: Forward 1, speed up 1?
Arnav: Yeah, because… Actually, we don't need the speed up, because, see, each point on that graph is one second, each point is one second, and all of them are the same (length?) we don't need a speed up.

Liam and Arnav referred back to the graphs, and then discussed the appropriateness of relevant ViMAP commands, "Forward <Step-size>" and "Speed-up <Change-in-Step-size>", for modeling the graphs. In doing so, they coordinated the change in distance as represented on the y-axis and time (x-axis) with the motion of an object, as Arnav's suggestion of a downward loop, gets taken up in joint action through *constructive listening* (Greeno & van de Sande, 2007) by Liam. Arnav then clarified that their ViMAP distance model did not require acceleration, because the length of the line (in the graph) during each second is the same, thereby implying that the object traveled the same distance during each second. This reflects coherence between an extrinsic graph perspective ("each point on that graph" and "all of them are the same (length?)"), an agent-perspective ("we don't need a speed up"), and an implied object perspective. For the first time in the activity, Arnav computationally parsed the difference between moving forward at the same speed and speeding up.

### Shifts in Point-of-View During the Interaction: A Timeline

Figure 3A illustrates the counts of points of view, coded line-by-line throughout the interaction, for both Liam and Arnav. Intermittent periods of unrelated, off-task are not included in this analysis. Since a focus of this analysis is on Arnav's changing definition of distance from the starting point, his point-of-view counts are separated from Liam's and shown in Figure 3B. Four perspectives are represented: the object perspective, the agent perspective, the perspectives of one another, and perspectives based on the graph. The perspectives are color coded in black, teal, green, and yellow, respectively. At the beginning of the interaction (0:00 - 2:51), both students primarily took a graph perspective. The segment reported in Episode 1 illustrates Arnav's extrinsic graph perspective, while Liam adopted an intrinsic graph perspective. As the interaction continued, the students focused on generating a program for the turtle so that it would carry out specific actions (2:52 - 04:43), then edited the program and reran it (4:44 - 7:29), in order to generate the shape of the graph. Here, three perspectives are at work simultaneously: the graph (both extrinsic and intrinsic), turtle (agent), and egocentric (i.e., each other).

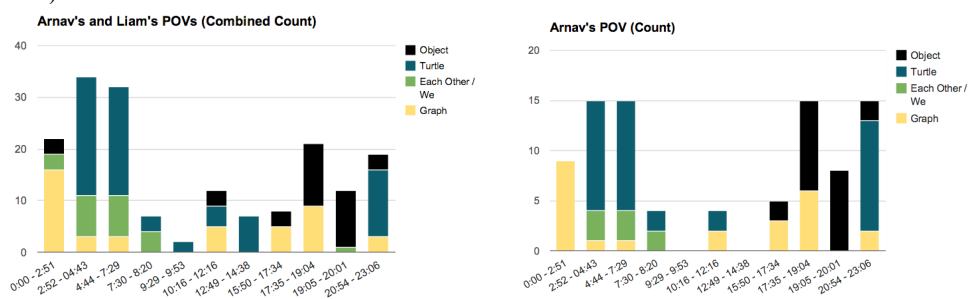

Figures 3A (left) and 3B (right). Arnav's & Liam's point-of-view, combined count (Figure 3A, left); and Arnav's points-of-view (Figure 3B, right).

Between 7:30 and 8:20, the instructor (second author) clarified the intended meaning of "distance" to the entire class, and Liam re-voiced this definition of distance to Arnav, who did not recognize any difference between that definition and the speed of the object. This was followed by an interaction between a facilitator and Liam, in which Liam explained the meaning of their (incorrect) model, but Arnav did not participate (09:29 - 09:53). In an extended interaction with the first author, (10:16 - 18:07), the students realized that their model showed changes in speed of the object in motion, but not its displacement. Episode 2, in which Arnav put forward a definition of "distance," began immediately after Amy left the students to change their model (18:09). In this segment, students integrated extrinsic and intrinsic graph perspectives with an object perspective (17:35 - 19:04). In Episode 3 (19:05 - 20:01 and 20:54 - 23:06), they begin enacting the motion (almost exclusively from an object perspective). However, when they began working on their program again, they coordinated the graph, agent perspective, and object perspectives (20:54 - 23:06). As these perspectives began to cohere, Arnav demonstrated his emerging understanding of the relationships among speed, distance traveled, and time, as shown in Episode 3.

## Conclusion and Discussion

We have argued that when students engage in collaborative agent-based programming in order to model motion as a process of change over time, the development of computational thinking and learning physics co-occur through students' negotiations of multiple perspectives or POVs. *Coherence* between these perspectives serves as the *constraint* (Thagard & Verbeurgt, 1998), and the students' understanding of the relevant physics is propelled forward through a process of constraint satisfaction as these perspectives cohere. In our study, bringing about this coherence, in turn, was deeply tied to the children's computational *doing*, including agent-based programming and reflective discourse. In addition, the instructors' prompts also pushed the children toward particular points of view. Similar to previous studies using agent-based modeling (Wilensky & Reisman, 2006), our study also shows that the agent-perspective can indeed play a productive role in understanding the relevant scientific concepts; however, we also show that this perspective needs to be negotiated with others for conceptual growth. These other perspectives included the children's egocentric perspectives, perspectives based on the graphs, and that of the (imagined) physical entity in motion. Achieving coherence between all these perspectives enabled the learners to bridge what is happening *now* (i.e., the instantaneous position and speed of the object in motion) with what has happened *until now* (i.e., previous changes in the object's position and speed) – a feat that is challenging for even college-level physics learners (McDermott et al., 1987).

## Acknowledgements

Funding from the National Science Foundation (NSF Early CAREER # 1150230) is gratefully acknowledged. Thanks to Rogers Hall, Rich Lehrer & Kevin Leander for insightful comments on earlier drafts.